\documentclass[10pt,letterpaper]{article}
\usepackage{opex3}
\usepackage{epstopdf}
\usepackage[]{textcomp}
\usepackage{amsmath}
\usepackage[]{xspace}
\usepackage{cite}
\begin{document}
\title{Toward real-time quantum imaging with a single pixel camera}
\author{B.J. Lawrie and R.C. Pooser$^{*}$}
\address{
Computational Sciences and Engineering Division, Oak Ridge National Laboratory, Oak Ridge, TN 37831 USA\\
{\color{blue}$^*$pooserrc@ornl.gov}
}

\begin{abstract}
We present a workbench for the study of real-time quantum imaging by measuring the frame-by-frame quantum noise reduction of multi-spatial-mode twin beams generated by four wave mixing in Rb vapor. Exploiting the multiple spatial modes of this squeezed light source, we utilize spatial light modulators to selectively pass macropixels of quantum correlated modes from each of the twin beams to a high quantum efficiency balanced detector. In low-light-level imaging applications, the ability to measure the quantum correlations between individual spatial modes and macropixels of spatial modes with a single pixel camera will facilitate compressive quantum imaging with sensitivity below the photon shot noise limit.
\end{abstract}
\ocis{(270.6570) Squeezed states; (270.0270) Quantum optics.}

\noindent 

\bibliographystyle{osajnl}

\section{Introduction}
Quantum imaging, which is based on the control of images in quantum systems, has been a topic of growing interest in recent years \cite{Boyer2008,Glorieux2012,Ding2012,Clark2012,Brambilla2008}. For an ideal laser source used in a classical imaging system with a mean number of photons $N$, the uncertainty in the number of photons is $\sqrt{N}$ and the Heisenberg  uncertainty principle dictates that the uncertainty in phase varies as $1/\sqrt{N}$.  This uncertainty corresponds to the photon shot noise limit (SNL), a fundamental result of the Poissonian statistics of coherent light sources.  With recent advances in the demonstration of highly multi-spatial-mode quantum light sources \cite{Boyer2008PRL,Corzo2012}, it is now possible to take advantage of quantum correlations between individual coherence areas in order to perform imaging with sensitivity or resolution \cite{Kolobov2000} beyond classical limits and to implement parallel quantum information protocols \cite{Lassen2007}.  In particular, quantum noise reduction (QNR) - or squeezed light - can be exploited in low-light-intensity imaging applications in order to achieve increased sensitivity and contrast beyond the SNL by reducing the uncertainty in photon number at the expense of the uncertainty in phase.  Other authors have recently demonstrated quantum imaging of individual images with sensitivity below the SNL with both discrete photon sources \cite{Brida2010} and continuous variable sources \cite{Clark2012}.   However, rather than examining the excess noise properties present in twin beams as a way to estimate the geometry of a quantum image \cite{Clark2012}, we demonstrate that direct analysis of squeezing can be used to the same end.

Because any attenuation of twin beams results in a degradation of quantum correlations \cite{Aytur1992,Smithey1992}, previous research demonstrating imaging with sub-shot-noise level sensitivity with CCD cameras relied on expensive high quantum-efficiency CCD arrays \cite{Brida2010}. The ability to perform sub-shot-noise imaging with a more economical, high quantum efficiency, single pixel detector with a dramatically reduced overall integration time will facilitate real-time high-sensitivity imaging in low-light situations.  In this manuscript, we demonstrate real-time control of the spatial modes in a squeezed light source by incorporating a spatial light modulator (SLM) into the seed beampath prior to the generation of multi-spatial-mode squeezed light by four-wave mixing (4WM) in  ${}^{85}$Rb vapor. This `quantum movie projector' exhibits real-time quantum noise reduction as a continuous series of images are flashed across the SLM, thereby allowing for the analysis of the quantum noise reduction present in arbitrary spatial modes in real-time.   There is a limited literature that applies compressive sampling algorithms to pseudothermal \cite{Jiying2010,Katz2009} and quantum \cite{Zerom2011} ghost  imaging, but there is currently no literature examining the high sensitivity imaging possible with the application of compressive sampling to squeezed light sources.  With the addition of two SLMs after the Rb vapor cell, we measure the quantum noise reduction between macropixels within the twin beams, a significant step towards real-time compressive quantum imaging with a single pixel camera.  In addition, we demonstrate compressive sampling of the beam profiles using a customized sampling matrix suited to twin beam detection.

\section{Experimental techniques}
The squeezed light source used in this experiment was generated by a 4WM process in  ${}^{85}$Rb vapor based on a double-$\Lambda$ system between the hyperfine ground states and the D1 excited states.  The hot ${}^{85}$Rb vapor in a 12.5 mm thick antireflection coated glass cell absorbs two photons from the pump beam (weakly focused to a waist of 1 mm and denoted by `P' in Fig. 1(b)) thereby generating a coherence between the two ground states. The probe beam (focused to a waist of 450 $\mu$m and denoted by `Pr' in Fig. 1(b)) is red-shifted from the pump frequency by roughly 3 GHz via a double-pass acousto-optic-modulator in order to stimulate the re-emission of photons into the probe frequency with the simultaneous emission of photons into the conjugate beam  (`C' in Fig.  1(b)) that is blue-shifted from the pump by 3 GHz.  Because of the strong amplitude correlations between these twin beams, the amplitude difference noise measured with a balanced photodiode and shown in Fig. 1(c) is below the photon shot noise limit (SNL) for most sideband frequencies between 50 kHz and 5 MHz.  All squeezing values reported in this manuscript were measured at a sideband frequency of 500 kHz with 20 kHz resolution bandwidth and 3 kHz video bandwidth.  In each case, the average of the 401 datapoints in each spectrum was used as the reported noise level. The combined systematic and statistical uncertainty in each data point is 0.1 dB.  Because of the presence of a buffer gas in the Rb vapor cell used for this experiment, increased Doppler broadening limited the maximum squeezing to 4.5 dB below the SNL.  Vapor cells with no buffer gas demonstrate 15-20\% less absorption of the probe field within the cell, yielding quantum noise reduction of greater than 9 dB inferred with no losses.

\begin{figure*}
\centerline{
\includegraphics[width=12 cm]{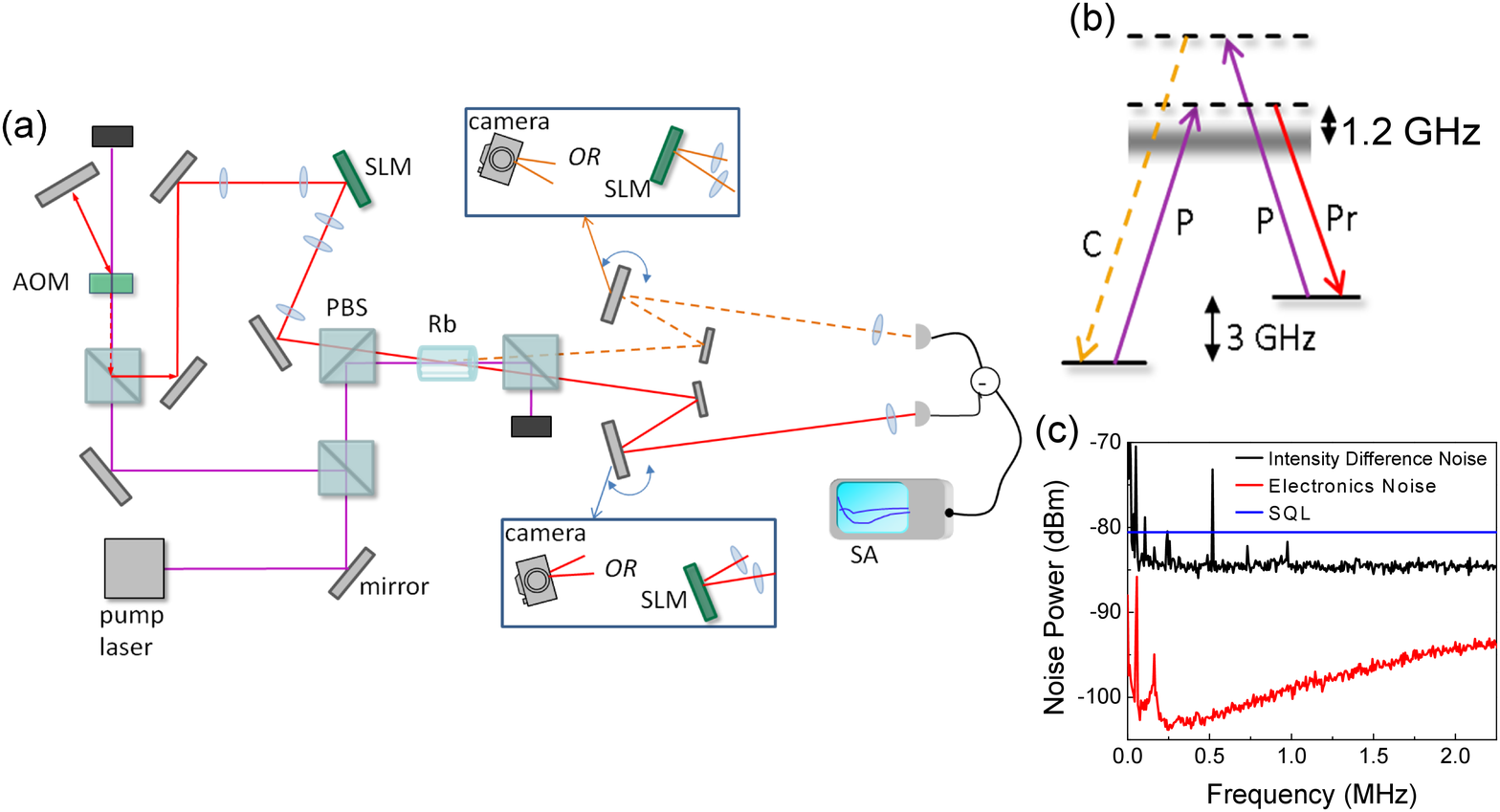}
}
\caption{(a) Schematic of the quantum imaging experiment that utilized a DMD as an SLM to control the spatial modes present in the twin beams and either two CCD cameras or two DMDs and a balanced photodiode to analyze the image quality and quantum noise reduction after 4WM.  (b) an energy diagram of the 4WM process at the D1 transition in  ${}^{85}$Rb, and (c) a typical squeezing spectrum demonstrating quantum noise reduction 4.5 dB below the SQL.}
\end{figure*}

A digital micromirror device (DMD) comprising a 1024 x 768 array of square micromirrors with a pitch of 13.68 $\mu$m provided intensity SLM functionality, thereby yielding real-time control over what spatial modes undergo four-wave mixing in the vapor cell.  Because the quantum noise reduction observed in individual coherence areas within an image is highly dependent on the pump-probe overlap, the observed squeezing is expected to be smaller for images with higher order spatial frequencies that would yield poor overlap at the Fourier plane in the vapor cell. This is a similar result to that seen when masks were used to imprint various images on the probe beam prior to four-wave mixing \cite{Boyer2008}.  In order to examine the quantum noise reduction in arbitrary images, two experimental designs were used.  First, the probe and conjugate were imaged with CCD cameras placed in the respective image planes, and flip mirrors were used as shown in Fig. 1(a) to measure the quantum noise reduction of each image with a balanced photodiode.  Subsequently, two additional DMDs replaced the cameras in the probe and conjugate image planes as shown in Fig. 1(a), and the DMDs were used to selectively pass macropixels of each beam to the photodiode in order to examine the feasibility of single pixel quantum imaging.

\begin{figure}[ht]
\centerline{
\includegraphics[width=9 cm]{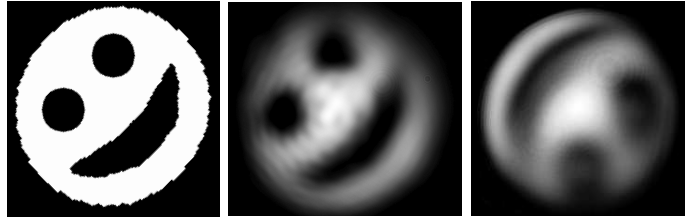}
}
\caption{(left) A bitmap image of a happy face 2 mm in diameter that was programmed onto the DMD spatially coincident with the incident probe beam, with the probe (middle) and conjugate (right) images acquired in the respective image planes after four-wave mixing.}
\end{figure}

\section{Real-time quantum imaging}
As an initial demonstration of the dependence of quantum noise reduction on the spatial modes present in the probe, various images were placed on the SLM illustrated schematically in Fig. 1(a) and imprinted on the probe seed beam, and the image quality and quantum noise reduction of the twin beams generated by 4WM were recorded.  For each image, the optics illustrated in Fig. 1 were unchanged so that while the probe waist remained centered in the vapor cell, the probe diameter in the cell varied with the structure of the Fourier image.  Compared with gaussian twin beams demonstrating 4.5 dB of quantum noise reduction, the happy face twin beams illustrated in Fig. 2 demonstrated 3.8 dB of quantum noise reduction below the SNL.  The spatial overlap between the pump and probe therefore effectively functions as a low-pass filter that removes higher frequency spatial modes from the Fourier image, resulting in reduced squeezing in images containing significant high-spatial-frequency components.  This is made more clear in Fig. 3, where the quantum noise reduction associated with several other images is presented: checkerboard patterns--possessing significant high frequency spatial modes--demonstrated the least squeezing.

\begin{figure}[hb]
\centerline{
\includegraphics[width=9 cm]{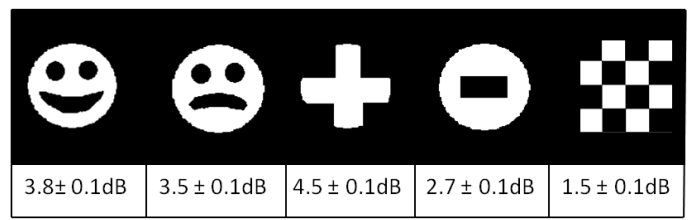}
}
\caption{The quantum noise reduction associated with various probe seed beam profiles.  The images in the top row were imprinted on the probe beam prior to 4WM.}
\end{figure}

In order to transition from the examination of quantum noise reduction in static images to real-time quantum imaging, a simple video of a spinning cross was implemented on the DMD prior to the Rb vapor cell, with the center of the cross coincident with the center of the gaussian probe beam.  The quantum noise reduction observed in real-time on a spectrum analyzer was unchanged from the quantum noise reduction of a static cross image.  Ringing in individual micromirrors resulted in the observation of noise peaks at isolated frequencies below 5 MHz within the QNR spectrum.  In order to eliminate any noise due to micromirror ringing effects, a 10 ms delay was introduced between the introduction of each image to the DMD and each QNR measurement.  A video of the probe and conjugate beam profiles recorded in the probe image plane--38 cm from the vapor cell is shown in Fig. 4(a)--with the corresponding real-time quantum noise reduction plotted as a function of cross angle in Fig. 4(b).  The probe and conjugate image planes are separated in the propagation direction due to cross phase modulation with the pump near resonance (the probe is near resonance while the conjugate is 6 GHz away). The buffer gas in the vapor cell caused pressure broadening of the absorption resonances, which increased the cross phase modulation between the probe and pump fields compared to a pure Rb cell. As a result, the probe and conjugate image planes were separated by 483 cm.  As can be seen in Fig. 4(a) the conjugate beam profile at the probe image plane is the Fourier transform of a spinning cross, while the video of the conjugate beam profile recorded in the conjugate image plane and shown in Fig. 4(c) is clearly that of a spinning cross.

\section{Single pixel
 quantum imaging}

While the videos in Fig. 4 are evidence of real-time quantum imaging, they were recorded asynchronously from the quantum noise reduction measurements, with the flip mirrors illustrated in Fig. 1(a) used to separately facilitate the recording of images or squeezing spectra.  Because the DMD shown in Fig. 1(a) provided us complete control over the images present in the quantum video, it was straightforward to demonstrate the reproducibility of these measurements, and it was clear which squeezing spectra corresponded to which images.  In the case of more arbitrary real-time quantum imaging applications, it will be necessary to synchronously measure both the quantum noise reduction and the spatial properties of the twin beams.  The use of SLMs that selectively pass either individual correlated coherence areas or macropixels of many coherence areas onto a balanced photodiode will allow for real-time quantum imaging with a single pixel camera, enabling myriad applications in low light, sub-shot-noise imaging.  The two DMDs in the probe and conjugate image planes illustrated in Fig. 1(a) facilitated the measurement of quantum noise reduction in macropixels of individual coherence areas--a significant step toward real-time single pixel quantum imaging with macroscopic photon numbers.  

\begin{figure}[t]
\centerline{
\includegraphics[width = 12 cm]{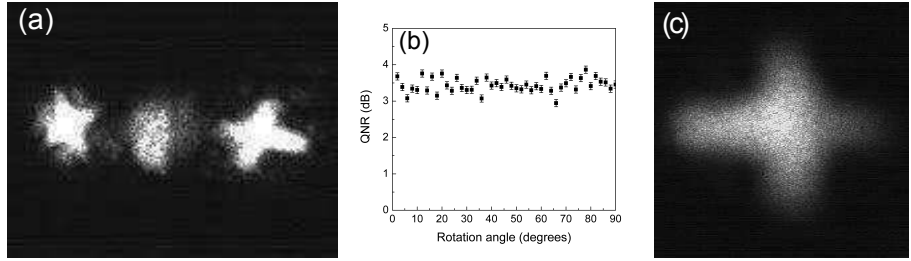}
}
\caption{(a) Video of conjugate, pump and probe beam profiles from left to right at the probe image plane, (b) quantum noise reduction as a function of cross rotation angle, and (c) a corresponding video of the conjugate in the conjugate image plane.}
\end{figure}

As a first demonstration of this capability, the probe beam profile was recorded by using the probe DMD and a photodiode as a rastered single pixel camera.  The left-most image in Fig. 5 is the thresholded image of the probe beam profile that was recorded in this manner.  When the thresholded image was placed on the probe DMD, and the conjugate DMD was set to pass all spatial modes, with a variable neutral density filter introduced to proportionately attenuate the conjugate--essentially acting as a bucket detector--1.67 dB of QNR was recorded.  The significant loss in squeezing compared with the 4.0 dB of QNR recorded with mirrors in place of the two DMDs is a result of the losses associated with the DMDs \cite{Aytur1992,Smithey1992}: when the DMDs reflect all spatial modes, 40\% attenuation is observed in the zero order diffraction spot.
\begin{figure}[h!]
\centerline{
\includegraphics[width= \columnwidth]{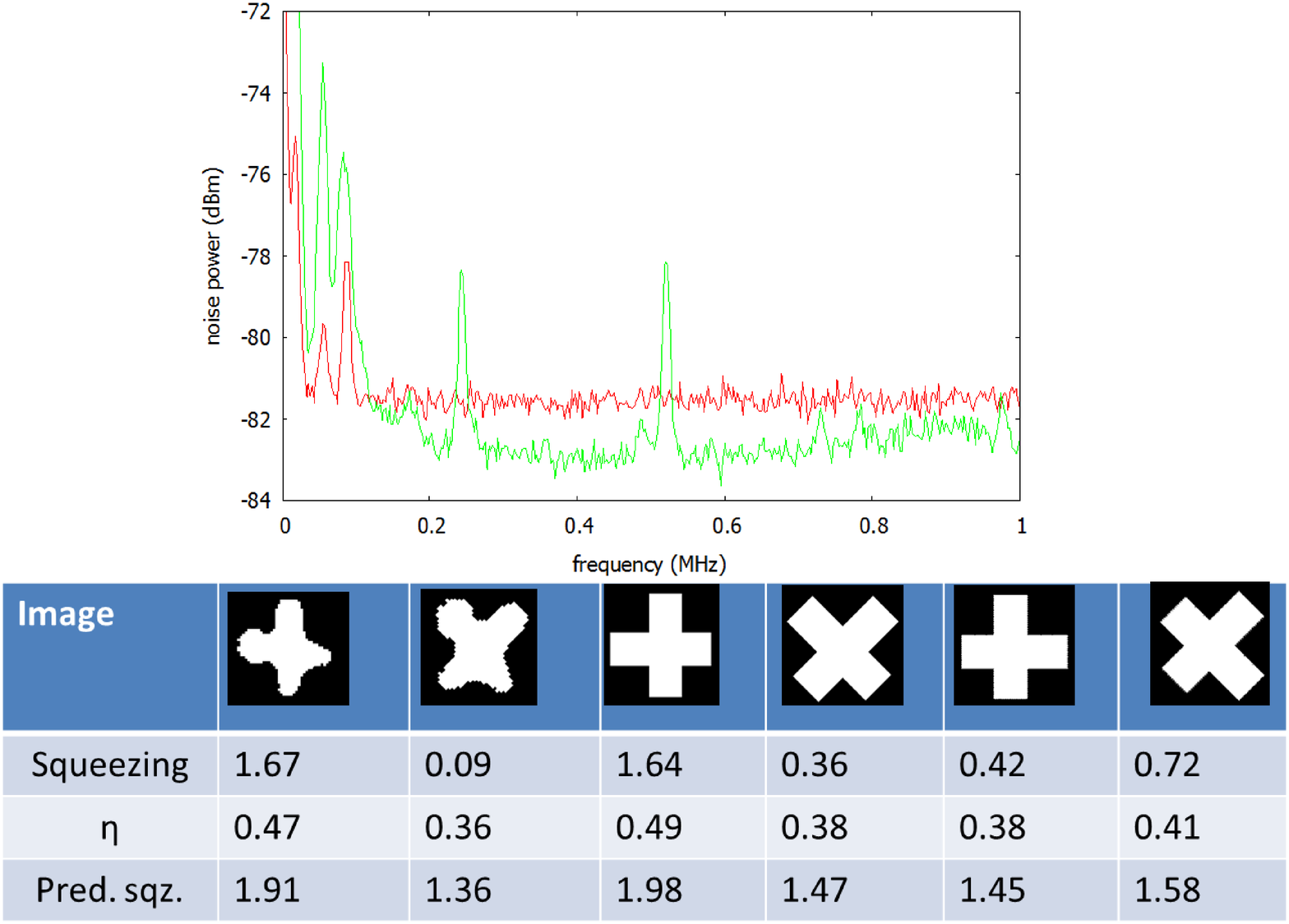}
}
\caption{(bottom) The quantum noise reduction associated with various masks introduced to the DMD, and (top) the quantum noise reduction spectrum associated with the leftmost bottom image. The total transmission ($\eta$) and predicted squeezing for a single spatial mode are shown for each image. The uncertainty associated with each squeezing value is 0.1 dB, while the uncertainty on each transmission measurement is 1\%.}
\end{figure}\label{fig5}

The predicted squeezing values in Fig.~5 were calculated using the input-output relations for a \textit{single spatial mode} phase-insensitive amplifier \cite{Smithey1992}: $S=10\log_{10}(1-\eta-\eta/G)$, where $G$ is the total gain of the 4WM process (approximately 4 in this experiment), and $\eta$ corresponds to the transmission of the vapor cell-DMD-detector system. While the DMD generally exhibits 60\% diffraction efficiency into the zero order mode, additional attenuation occurs with each image placed on its surface, since some incident light falls onto pixels which are turned ``off'' (set to reflect light away from the detector).

As seen in the bottom of Fig. 5, slight variations in the image present on the probe DMD resulted in dramatic changes to the observed QNR.  For each image, the conjugate DMD passed all spatial modes and the conjugate power was proportionately attenuated in order to eliminate excess noise and optimize squeezing. In all cases the conjugate attenuation that resulted in optimal squeezing was equal to the probe attenuation within 1\%. Because the predicted squeezing was provided for a single spatial mode, the losses are applied to the entire beam equally rather than to individual coherence areas. For certain images the measured squeezing is much lower because non correlated coherence areas between the probe and conjugate are being detected. This provides a novel technique for quantum imaging analogous to a recently reported technique that relied on measurements of excess noise made by homodyne detection to provide rudimentary quantum imaging without the use of a CCD camera \cite{Clark2012}.  In contrast to that manuscript, this technique allows us to approximate an object's size and shape via the direct examination of quantum noise reduction as a function of input image in the SLM.  The strong dependence of the squeezing on the geometry present on the probe DMD is a result of the multi-spatial-mode nature of the quantum correlations present in the twin beams.

In order to progress to true single pixel quantum imaging, a simple solution is to selectively pass quantum correlated coherence areas with each DMD, resulting in a rastered single pixel quantum image.  Such techniques are slow, and acquiring the images with only one super pixel at a time would likely introduce sufficient attenuation to make the subsequent observation of squeezing very difficult.   Compressive imaging resolves this issue with the utilization of sampling matrices containing many pixels.  As a first step toward single pixel quantum imaging, preliminary rastered quantum imaging results were obtained by rastering a line instead of individual pixels.  The line illustrated in  Fig. 6(a) was centered on the location of a gaussian conjugate beam on the conjugate DMD in order to pass only those conjugate photons that were spatially coincident with the line.  A line scaled in width by the ratio of the probe and conjugate image sizes at their respective image planes was rastered across the probe DMD as shown in Fig. 6(b). Excess noise was observed for all line positions until they sufficiently overlapped corresponding areas in the probe and conjugate beams to make squeezing observable.  The quantum noise reduction spectrum shown in Fig. 6(c), recorded when both lines were centered on the twin beams, shows 1$\pm$0.1 dB squeezing at 500 kHz, indicating that this approach should be sufficient to allow for rastered quantum imaging.  Correspondingly, these results indicate the feasibility of compressive quantum imaging when a sampling matrix having a comparable number and arrangement of pixels is selected. In order to demonstrate the effect of a reduction in macropixel size on the measured squeezing, this experiment was repeated with lines having half the width of the lines illustrated in Figs. 6(a) and 6(b) (corresponding to approximately one fourth of the $e^{-2}$ beam diameter at the DMD).  In this instance, a line was centered on the probe beam, and a separate line was rastered across the conjugate DMD.  As illustrated in Fig. 6(d), excess noise was observed for all beam displacements except where the line was centered on the conjugate beam.  The goal in this experiment was to show that images could be placed on the probe and conjugate DMDs simultaneously in correlated locations such that squeezing was still observable. This is important to demonstrate because the types of sampling matrices used in our compressive imaging algorithm correspond to shapes almost exactly like those shown in the bar image. Figure 7 shows an example of one such sampling matrix.  We have also perfomed initial compressive imaging with sampling matrices of this form. 

\begin{figure}
\centerline{
\includegraphics[width= \columnwidth]{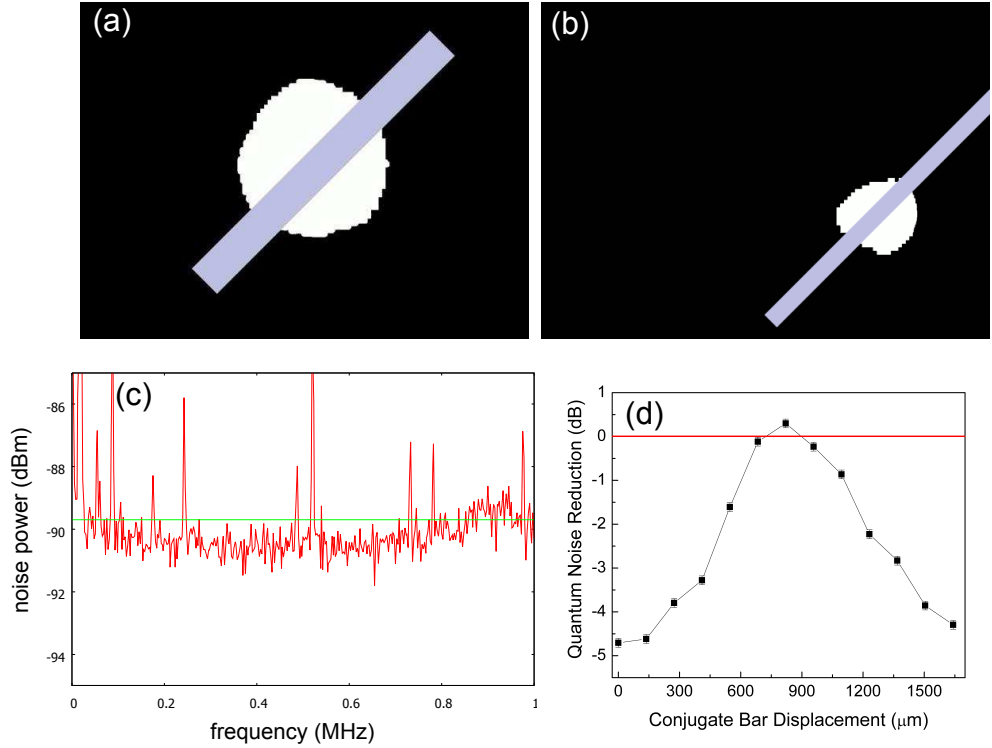}
}
\caption{Thresholded images of the conjugate (a) and probe (b) gaussian beam profiles in white.  The line on the conjugate beam profile was placed on the conjugate DMD in the center of the conjugate beam profile.  The line on the probe beam profile was rastered across the probe on the probe DMD.  The QNR spectrum (c) shows the quantum noise reduction of approximately 1 dB at 500 kHz that emerged when the mask was centered on the probe beam profile.  The reduction of the width of the line by a factor of 2 yielded a dramatic reduction of squeezing even in the case where corresponding areas of the probe and conjugate were passed by the respective DMDs (d).}
\end{figure}

An important question is the relation between super pixel size and coherence area. In simple terms, a coherence area acts as a single coherent beam. If a single pair of correlated coherence areas in the two beams were isolated and studied, one would find that the quantum noise reduction as a function of transmission is effectively independent of the spatial structure of the attenuator. That is, uniformly attenuating each beam would yield the same results as attenuating them with sharp apertures as long as each beam consisted of only one coherence area. On the other hand, if the beams contain multiple coherence areas, then the noise reduction as a function of transmission would depend strongly on the spatial structure of the attenuation. This effect is illustrated in Fig.~5. Brida \textit{et al.} \cite{Brida2010} found that experimentally maximizing the ratio of super pixel size to coherence area leads to reduced noise, which agrees with the theoretical findings of Brambilla \textit{et al.} \cite{Brambilla2008} The coherence area size is estimated from the conjugate size inside the vapor cell and the beam size in the far field to be 1.4 mrad (about 17\% larger than in reference \cite{Boyer2008}), while the angular beam diameter of the probe and conjugate was measured to be 2.4 mrad. Therefore approximately three to four coherence areas were present within the $1/e^2$ diameters of the probe and conjugate beams. Note that this is the number of coherence areas within the total angular bandwidth of the 4WM process that the nonzero intensity regions of the probe and conjugate overlap with, and the low intensity tails of the probe and conjugate contain yet more coherence areas beyond those located within the Gaussain beam width. The actual number of supported modes within the angular bandwidth of the four wave mixing process was found to be approximately 70 using a previously reported method\cite{Boyer2008PRL}. The bar sizes were chosen to be approximately equal to half the angular diameter ($1/e^2$ radius) of each beam, which suggests that the bars used as macropixels in Fig.~6(a) and 6(b) likely contained only individual coherence areas in the narrow dimension. Thus, the bar size chosen is likely close to the smallest size allowable to observe squeezing. However, the bar size can be smaller and still exhibit quantum noise reduction as a result of subsampling individual, corresponding coherence areas in each beam (this case would look like simple attenuation for a single mode, and noise reduction would reduce to zero with complete attenuation). The exact distribution of the coherence areas within the beams likely does not correspond to a striped shape (but rather would correspond to the shape of the Schmidt modes). Thus, the dramatic reduction in squeezing resulting from the choice of narrow bar sizes and shown in Fig. 6(d) is a result of subsampling individual coherence areas while introducing excess noise from uncorrelated coherence areas. We note that determining the exact shape of the coherence areas may be possible using the DMDs employed in this experiment by selectively rastering each beam with specific shapes and sizes for superpixels, but such a study is beyond the scope of the present work.

We can also determine the average power per coherence area illuminated by the probe and conjugate by multiplying the total beam power by the ratio of the coherence area size to total beam size. The probe in Fig.~6 had approximately 1.2 mW total power, corresponding to approximately 400 $\mu$W per coherence area. The actual number is slightly less, because the nonzero intensity size of the beams on the DMD extended to approximately three to four times the beam width, meaning the total power was distributed among slightly more coherence areas. Nonetheless, this power per coherence area is in stark contrast to quantum imaging experiments with spontaneous parametric downconversion which rely on photon counting, and easily satisfies the condition of a large number of photons per coherence area needed for quantum imaging.

Primary issues with performing rastered single pixel imaging include long acquisition times and low signal to noise ratios associated with small pixel sizes.  Both of these issues can be addressed with the increasingly popular technique of compressive imaging (CI).  CI deals with the acquisition of images by recovering the coefficients in a sparse representation basis.  An image containing N pixels requires only M$\ll$N coefficients to provide a good approximation of the image x under the transform:
\begin{equation}
x=\Psi \alpha
 \label{eq:CI
}
\end{equation}
where $\Psi$ is the sparsifying basis and $\alpha$ is the coefficient vector comprising no more than K significant nonzero values.  CI theory provides the framework to acquire such an image via the linear projection:

\begin{equation}
y=A x + \Gamma
 \label{eq:CIreconstruct
}
\end{equation}
where y is an M-dimensional measurement vector, A is an M-by-N sensing matrix, x is the N dimensional signal of interest, and $\Gamma$ is the acquisition noise.  CI is attractive for single pixel imaging applications because it allows for image reconstruction with only $M \approx K log N$ measurements, and the use of sparse arrays instead of individual pixels significantly increase the signal to noise ratio in applications where the dominant noise source is the shot noise.  In addition, real-time compressive imaging is now a viable technology \cite{Do2012,smith2012}. 

Optimization of CI for a given application requires two separate thrusts: the choice of appropriate sparse sampling matrices, and the choice of an appropriate reconstruction algorithm to facilitate the reconstruction of $x$ from $y$.  After comparing various permutations of random orthonormal matrices and Hadamard matrices it was determined that individual rows randomly sampled from block Hadamard matrices \cite{Do2012} modified using semilocal randomizers yielded an efficient and high quality image reconstruction \cite{pooserforthcoming}.  Reconstruction techniques including total variation with equality constraints from L1 magic \cite{candes2006}, gradient projection \cite{Figueiredo2007}, and total variation with augmented Lagrangian \cite{li2012} were investigated for this application.  Ultimately, the total variation with equality constraints reconstruction algorithm was found to reliably reconstruct beam profiles with optimal signal to noise ratios.  Figure 7 illustrates the 32x32 reconstructions of the probe and conjugate beam profiles generated from M=300 measurements by the total variation algorithm for a test image of the letter 'E'.  With no probe or conjugate DMD, the 'E' twin beams demonstrated 3.1$\pm$0.1 dB of squeezing.  By choosing sampling matrices with appropriate pixel sizes in comparison to the probe and conjugate coherence areas, similar to the demonstration in Fig. 6, it is possible to perform differential compressive sampling in order to acheive real-time single pixel compressive quantum imaging.
\begin{figure}
\centerline{
\includegraphics[width= 9 cm]{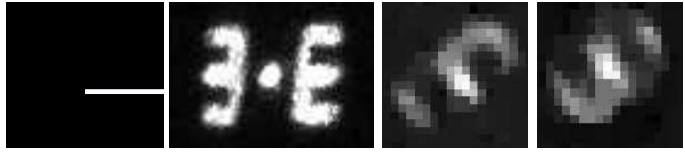}
}
\caption{An example sampling matrix used in the compressive imaging algorithm (left); beam profiles of probe and conjugate 'E' in probe image plane acquired with a CCD camera (middle left), beam profile of probe acquired with total variation minimization with equality constraints utilizing the probe DMD at the probe image plane (middle right), and the beam profile of the conjugate (right) acquired with compressive imaging techniques with the conjugate DMD placed at the conjugate image plane.}
\end{figure}

It is interesting to compare compressive imaging with other imaging methods such as rastering with CCD cameras or direct imaging using local oscillators. A big advantage for compressive imaging can be seen in the use of a large number of pixels to obtain each data point, ratehr than a single pixel per measurement. Fig.~7 shows one type of sampling matrix which sends light to the detector whenever light is incident on the ``white'' part of the DMD. A single pixel would correspond to a square approximately sixteen times smaller than the sampling matrix shown here. Further, the sampling matrices used in this experiment are scalable in size by selecting a block size for the block diagonal sparse array that generates the sampling matrices. This allows the signal to noise ratio of each sample to be scaled as needed to overcome detector dark noise, for instance. Thus, compressive imaging can offer in principle better signal to noise ratios compared to CCD cameras for a given quantum efficiency and dark current noise. Compared to the homodyne detection method of imaging, compressive imaging offers in principle the same resolution and signal to noise ratio for a given transmission, but offers no phase information. This is because a direct intensity measurement as performed in our experiments is analagous to a perfectly aligned homodyne detector with local oscillator phase locked to detect the amplitude quadrature. However, the ease of alignement using DMD's is a potential benefit over the difficulty of aligning a homodyne detector for maximum visbility, especially in more comon enironments which are not necessarily suited to interferometry. Deriving a local oscillator can in itself be difficult task as well. While compressive imaging certainly has drawbacks, its advtantages make it a good candidate for quantum imaging.

\section{Conclusions}
In this manuscript we have demonstrated all the necessary ingredients to achieve real time compressive quantum imaging. This will facilitate the development of high sensitivity low-light-intensity imaging systems that will prove valuable in applications ranging from covert imaging to the imaging of photosensitive biological samples.  The examination of QNR as the mask on a DMD was varied provides a novel technique for quantum imaging in addition to motivating the study of single pixel quantum imaging.   The rastered quantum imaging results of Fig. 6 make it clear that the appropriate choice of sampling matrices, in conjunction with the use of high-throughput SLMs--now commercially available with throughput greater than 95\%--will be sufficient to achieve real-time quantum compressive imaging with squeezed light for the first time.

\section*{Acknowledgments}

This work was performed at Oak Ridge National Laboratory, operated by UT-Battelle for the U.S. Department of energy under contract no. DE-AC05-00OR22725.  B.J.L was supported by a fellowship from the IC postdoctoral research program. R.C.P. acknowledges partial support from the Wigner fellowship.

\end{document}